\documentclass[12pt,a4paper]{article}
\usepackage[utf8]{inputenc} 
\usepackage[T1]{fontenc}    
\usepackage[english]{babel} 
\usepackage{mathptmx}  
\usepackage{amsmath,amssymb,amsthm}
\usepackage{mathtools}             
\usepackage{bm}   
\usepackage{booktabs}   
\usepackage{array}    
\usepackage{multirow}  
\usepackage{graphicx} 
\usepackage{adjustbox}  
\usepackage{caption}  
\usepackage[margin=2.5cm]{geometry} 
\usepackage{setspace}              
\usepackage{microtype}             
\usepackage{ragged2e}              

\usepackage{natbib}                
\usepackage{hyperref}
\hypersetup{
  colorlinks=true,
  linkcolor=blue,
  citecolor=blue,
  urlcolor=cyan
}

\newcolumntype{L}[1]{>{\RaggedRight\arraybackslash}p{#1}}

\title{\textbf{An Analysis of Monetary Policy Evidence and Theory through Meta-Analyses}}
\author{Ricardo Alonzo Fernández Salguero}
\date{\today}

\begin{document}

\maketitle

\begin{abstract}
\justify
This paper offers a synthesis of the empirical literature on the effects of monetary policy. Using the findings from an extensive collection of meta-analyses, it evaluates the effectiveness of conventional and unconventional monetary policy instruments on key macroeconomic variables such as output, inflation, capital flows, and the exchange rate. The aggregated evidence reveals a systematic gap between the effects reported in primary studies and the actual magnitude of these effects, once corrected for publication bias and methodological heterogeneity. The findings suggest that, while monetary policy is a relevant tool, its power to modulate the business cycle has been consistently overestimated in the literature. Contextual factors—such as the degree of financial development, the exchange rate regime, central bank independence, and crisis conditions—that modulate the transmission of monetary policy are identified. In particular, it is found that publication bias systematically favors statistically significant results consistent with predominant theory, which artificially inflates the perception of effectiveness. By correcting these distortions, a picture of monetary policy emerges with more modest, uncertain effects and considerable lags, which has profound implications for macroeconomic theory and the practice of economic policy.
\end{abstract}


\section{Introduction}
\justify
Monetary policy constitutes one of the most potent and debated tools in the arsenal of economic policy. Its ability to influence aggregate demand, stabilize inflation, and moderate the business cycle is a fundamental pillar of modern macroeconomics. However, despite decades of empirical research, a notable lack of consensus persists regarding the magnitude, speed, and stability of its effects. The literature is replete with estimates that vary considerably across countries, time periods, and econometric methodologies, generating a fragmented and, at times, contradictory landscape. As \citet{DeGrauwe2004} point out, the variance in empirical results is so large that it makes it difficult to formulate robust conclusions about policy effectiveness.

This heterogeneity is not a mere academic artifact; it has direct implications for central banks, which must make decisions in an environment of uncertainty. The question of whether a 100-basis-point increase in the benchmark interest rate will cause a modest or severe contraction in output, or whether it will cool inflation quickly or gradually, remains an empirical challenge. The proliferation of studies, far from converging towards a consensus, has often amplified uncertainty.

In this context, meta-analysis emerges as an indispensable tool. Instead of presenting one more primary study, this methodology allows for a quantitative synthesis of all available evidence, transforming a collection of individual studies into a single coherent body of knowledge. Meta-analysis not only averages the reported effects but also investigates the sources of heterogeneity among studies. It allows for identifying whether differences in results are due to systematic factors such as the methodology employed (e.g., VAR vs. GMM), data characteristics (panel data vs. time series), institutional context (developed vs. emerging economies), or, more worryingly, to biases inherent in the academic publication process.

This paper undertakes an analysis of the monetary policy literature through the lens of multiple meta-analyses. The objective is threefold: first, to establish the best estimate of the effect of monetary policy on output, inflation, and other key variables, once biases are considered and corrected. Second, to identify the systematic moderators that explain why effects vary so much across studies. Third, to critically evaluate the state of accumulated knowledge, contrasting conventional wisdom with the aggregated evidence emerging from quantitative synthesis.

To this end, a wide range of meta-analyses covering various facets of monetary policy are examined. Conventional monetary policy is analyzed, where works such as those by \citet{Enzinger2025} and \citet{Balima2020a} question the magnitude of its effects on output and inflation. Specific transmission channels are explored, such as interest rate pass-through \citep{Gregor2021}, money demand \citep{Knell2005,Knell2003,Kumar2014}, and exchange rate pass-through \citep{Velickovski2011, Iorngurum2025a}. Unconventional policy is addressed, analyzing its effects on the real economy \citep{Papadamou2019} and its international spillover effects \citep{Araujo2025}. Finally, it is investigated how structural factors such as central bank independence \citep{Iwasaki2017}, dollarization \citep{Korab2023}, or public debt levels \citep{Heimberger2023} modulate policy effectiveness. Through this synthesis, the aim is to offer a more nuanced and robust verdict on what we truly know about the power of monetary policy.

\section{The Contamination of Evidence: Publication Bias and Methodological Heterogeneity}
\justify
One of the most consistent and concerning findings revealed by meta-analyses in economics is the existence of profound publication bias. This phenomenon, also known as the file drawer problem \citep{Rosenthal1979}, refers to the tendency of researchers, reviewers, and editors to favor the publication of results that are statistically significant and/or consistent with predominant theory, while null or counter-intuitive results tend to remain unpublished. This bias systematically distorts the available body of evidence, creating a false impression of consensus and exaggerating the magnitude of real effects.

In the field of monetary policy, this distortion is particularly serious. The meta-analysis by \citet{Enzinger2025} on conventional monetary policy is perhaps the most forceful study in this regard. Based on a monumental database of 409 primary studies and over 146,000 estimates, the authors document an anomalous accumulation of results just above statistical significance thresholds (a fingerprint of p-hacking) and a marked asymmetry in funnel plots. Their verdict is unequivocal: the literature on the effects of conventional monetary policy is systematically biased towards publishing larger negative effects than the full evidence would warrant. \citet{Balima2020a}, in their meta-analysis on inflation targeting, reach a similar conclusion, finding a bias that favors the beneficial effects of this regime on inflation and GDP volatility. This bias is not limited to conventional policy; \citet{Nguyen2021} also detect it in the literature on the effects of monetary policy on prices in emerging economies, and \citet{Iorngurum2025a} corroborates it for exchange rate pass-through.

The direct consequence of this bias is an overestimation of monetary policy effectiveness. When \citet{Enzinger2025} apply correction techniques, the average peak effect of a 100 bps hike in the interest rate on output is reduced by half or more, from -1.0\% to a modest -0.25\%. Similarly, the effect on prices is reduced from -0.75\% to an almost imperceptible -0.15\%. These corrected effects are not only smaller, but their confidence intervals often include zero, casting doubt on their statistical significance. Table \ref{tab:correccion_sesgo} summarizes these findings, illustrating the gap between perceived and corrected effectiveness.

\begin{table}[h!]
  \centering
  \caption{Effects of Contractionary Monetary Policy (100 bps) Before and After Publication Bias Correction}
  \label{tab:correccion_sesgo}
  \begin{adjustbox}{width=\textwidth}
    \begin{tabular}{L{3.5cm} L{4.5cm} L{4.5cm} L{3.5cm}}
      \toprule
      \textbf{Variable} & \textbf{Reported Peak Effect (Uncorrected Average)} & \textbf{Corrected Peak Effect (Bias-Adjusted Average)} & \textbf{Main Source} \\
      \midrule
      Output & Maximum fall of approx. -1.0\% after 2 years. & Maximum fall of approx. -0.25\% after 1-2 years. & \citet{Enzinger2025} \\
      Price Level & Maximum fall of approx. -0.75\% after 4-5 years. & Maximum fall of approx. -0.15\% after 4-5 years. & \citet{Enzinger2025} \\
      Inflation Volatility & Significant reduction reported in the literature. & Genuine effect not significant after correction. & \citet{Balima2020a} \\
      GDP Growth & Null or slightly negative effect. & Genuine positive but not significant effect. & \citet{Balima2020a} \\
      \bottomrule
    \end{tabular}
  \end{adjustbox}
  \justify
  \footnotesize{\textit{Note:} The table synthesizes the findings of key meta-analyses that apply corrections for publication bias. Corrected effects are consistently of smaller magnitude than simple averages from the literature, suggesting a systematic overestimation.}
\end{table}

Beyond publication bias, heterogeneity in results is also due to methodological choices. \citet{DeGrauwe2004} were among the first to point out that the econometric technique used (VAR, SVAR, or structural models) is a significant predictor of the magnitude of effects. More recent studies confirm this dependence. For example, \citet{Iorngurum2025b} in their meta-analysis on interest rate pass-through, finds that model choice (static vs. dynamic, VECM, etc.) is a key factor in explaining heterogeneity. Similarly, \citet{Korab2023} show that the way dollarization is measured (de jure, de facto, deposits, etc.) and the estimation method systematically affect results on growth and inflation. \citet{Fidrmuc2020}, analyzing the New Keynesian Phillips Curve, also conclude that methodological and data characteristics have a significant impact on the estimated weight of the forward-looking component. In summary, empirical evidence is not a monolith; it is a mosaic of results strongly conditioned by researchers' decisions and the pressures of the academic publication system. Meta-analyses not only reveal this pattern but offer the tools to correct it and approach a more robust understanding of economic reality.

\section{Effects of Conventional Monetary Policy on Output and Inflation}
\justify
Conventional macroeconomic wisdom, formalized in New Keynesian models, posits that a contractionary monetary policy—typically an unanticipated hike in the benchmark interest rate—should reduce both output and inflation, albeit with lags. Output responds with a hump shape, falling to a trough after several quarters and then gradually converging to its equilibrium level, while the price level adjusts more slowly and permanently downwards. Meta-analyses of the empirical literature, as a first approximation, confirm this qualitative pattern. \citet{DeGrauwe2004}, in one of the first attempts at quantitative synthesis, found that a 100 bps increase in the interest rate reduced output by an average of -0.33\% after one year. However, their analysis did not employ the bias correction techniques that are standard today.

The work by \citet{Enzinger2025} offers the most comprehensive evaluation to date. As mentioned, the simple average of the 4,871 impulse response functions they analyze suggests a peak fall in output of -1\% and in the price level of -0.75\%. These values, though consistent with theory, hide the systematic distortion of publication bias. Once multiple correction methods are applied, the picture changes dramatically. The corrected peak effect on output is a modest -0.25\%, and on prices, -0.15\%. This massive reduction in magnitude has profound implications.

One of the most important is the \textit{sacrifice ratio}, defined as the cost in terms of lost output to achieve a one-percentage-point reduction in inflation. With the uncorrected data from \citet{Enzinger2025}, the sacrifice ratio (calculated as the quotient of peak effects) is approximately 1.3 (1.0/0.75). After correction, this ratio drastically worsens, standing around 1.7 (0.25/0.15). This suggests that, if the corrected effects are accurate, reducing inflation is substantially more costly in terms of economic activity than the uncorrected literature would imply. This finding undermines the narrative of an omnipotent central bank capable of managing the economy with precision and low costs, and brings us closer to the monetarist view of a crude tool whose effects are, at best, modest and uncertain.

The meta-analysis by \citet{Balima2020b} on inflation targeting adds another layer of complexity. While their analysis focuses on regime comparison, their findings on genuine effects are revealing. After correcting for bias, they find that inflation targeting is associated with lower inflation and higher GDP growth, but has no significant genuine effect on inflation volatility or GDP growth volatility. This challenges the idea that the primary benefit of inflation targeting is stabilization.

Table \ref{tab:efectos_convencional} summarizes and compares the findings of several meta-analyses on the effects of conventional monetary policy. The general conclusion is robust: the empirical literature, as a whole, has systematically overestimated the power of conventional monetary policy.

\begin{table}[h!]
  \centering
  \caption{Synthesis of Meta-Analysis Findings on Conventional Monetary Policy}
  \label{tab:efectos_convencional}
  \begin{adjustbox}{width=\textwidth}
    \begin{tabular}{L{4cm} L{3cm} L{4cm} L{5cm}}
      \toprule
      \textbf{Study} & \textbf{No. of Primary Studies} & \textbf{Peak Effect on Output (Corrected)} & \textbf{Key Findings on Biases and Heterogeneity} \\
      \midrule
      \citet{DeGrauwe2004} & 43 & Uncorrected (approx. -0.33\% per year) & Methodology (VAR vs. SVAR) and inflation level are key determinants. No formal publication bias tests performed. \\
      \citet{Balima2020b} & 113 (8059 estimates) & Higher GDP growth under IT (genuine effect) & Strong publication bias favoring beneficial effects of IT. Actual impact varies by country-specific factors and implementation. \\
      \citet{Enzinger2025} & 409 (4871 IRFs) & Fall of approx. -0.25\% & Robust and quantitatively very important publication bias and p-hacking. Shock identification choices are less important than bias. \\
      \citet{Nguyen2021} & 45 & Not applicable (focused on prices) & Publication bias detected. Effects on prices are weaker in countries with fixed exchange rates and greater financial development. \\
      \bottomrule
    \end{tabular}
  \end{adjustbox}
  \justify
  \footnotesize{\textit{Note:} The table illustrates the evolution and convergence of meta-analysis findings. More recent studies, with larger databases and more sophisticated correction techniques, point to smaller effects and a greater importance of publication bias.}
\end{table}

\section{Transmission through Specific Channels: Interest Rates, Money Demand, and Exchange Rates}
\justify
The effectiveness of monetary policy depends on the strength and reliability of its transmission channels. Meta-analyses have been useful in evaluating the empirical evidence on these mechanisms, often revealing them to be weaker and more heterogeneous than simple theory would suggest.

\subsection{Interest Rate Pass-Through}
\justify
The interest rate channel is the quintessential transmission mechanism. A change in the policy rate is expected to pass through to money market rates and, subsequently, to bank lending rates for businesses and households. The meta-analysis by \citet{Gregor2021} offers the most comprehensive synthesis of this literature. Based on 1098 estimates from 54 studies, they find that pass-through is, on average, incomplete. The average pass-through coefficient is approximately 0.8, meaning that only 80\% of a change in the benchmark rate is transmitted to lending rates in the long run.

Furthermore, the study reveals systematic heterogeneity. Pass-through is significantly lower for consumer loans and long-term loans compared to short-term corporate loans. This implies that monetary policy has more difficulty influencing household spending decisions and long-term investment. A particularly interesting finding is the impact of the Global Financial Crisis (GFC) of 2008. After the crisis, pass-through generally weakened, an effect the authors attribute to greater trade openness, higher volatility, and lower central bank independence in the post-crisis period.

\subsection{Money Demand and its Income Elasticity}
\justify
The stability of money demand is a necessary condition for monetary aggregates to serve as a reliable intermediate target for monetary policy. The empirical literature on the income elasticity of money demand is vast, and its results are highly varied. Several meta-analyses have tried to bring order to this evidence. \citet{Knell2003, Knell2005} were pioneers in this field. Their work, based on nearly 500 estimates, shows that income elasticity systematically depends on the definition of the monetary aggregate (higher for broad aggregates like M2/M3 than for narrow aggregates like M1), the inclusion of wealth as an explanatory variable (which reduces the estimated income elasticity), and financial innovation (which also reduces it). A key finding is the difference between countries: income elasticity is significantly lower in the U.S. than in other OECD countries.

\citet{Kumar2014} extends this work by explicitly comparing advanced and developing countries. They confirm that for advanced countries, broader monetary aggregates have higher income elasticities. However, for developing countries, this difference is only marginal. This suggests that portfolio decisions driving demand for broad money are more relevant in economies with more sophisticated financial systems. Similarly, wealth and financial reforms have a significant impact on reducing income elasticity in advanced countries, but a weak or insignificant effect in developing countries. \citet{El-Shagi2022}, in a specific meta-analysis for China, confirms the importance of the institutional context, finding strong suggestion of stable long-run money demand once publication bias is controlled for.

\subsection{Exchange Rate Pass-Through}
\justify
In open economies, the exchange rate is a transmission channel. The extent to which exchange rate fluctuations are passed through to domestic prices is known as exchange rate pass-through (ERPT). High ERPT implies that monetary policy can have rapid inflationary effects through currency depreciation, which often leads to a fear of floating. \citet{Velickovski2011} conducted a meta-analysis comparing transition and developed economies. Based on 575 coefficients from 23 studies, they confirm that ERPT is incomplete. Their most important results refer to the differences between the two groups of countries. They find no statistically significant difference in ERPT to import prices. However, ERPT to consumer prices is \textit{significantly and substantially higher} in transition economies than in developed economies. This finding provides a solid empirical basis for the reluctance of many monetary authorities in transition economies to allow greater exchange rate flexibility.

\citet{Iorngurum2025b}, in a more recent and extensive meta-analysis covering 1219 estimates for 111 countries, delves into the sources of heterogeneity. Using Bayesian Model Averaging (BMA), they find that results vary due to a combination of country-specific and methodological factors. Country-specific factors include trade openness, exchange rate flexibility, economic development status, and exchange rate persistence. Methodological factors include estimation methods, data characteristics, and endogeneity bias. One finding is asymmetry: a 1\% increase (depreciation) in the exchange rate leads to a 0.19\% decrease in the consumer price level, while a 1\% decrease (appreciation) leads to a decrease of only 0.09\%. Table \ref{tab:canales_transmision} synthesizes these findings.

\begin{table}[h!]
  \centering
  \caption{Synthesis of Meta-Analyses on Specific Transmission Channels}
  \label{tab:canales_transmision}
  \begin{adjustbox}{width=\textwidth}
    \begin{tabular}{L{3.5cm} L{3.5cm} L{9cm}}
      \toprule
      \textbf{Transmission Channel} & \textbf{Main Study} & \textbf{Key Findings} \\
      \midrule
      Interest Rate Pass-Through & \citet{Gregor2021} & Pass-through is incomplete (approx. 0.8). It is weaker for consumer and long-term loans. Weakened after the GFC due to macro-financial factors and lower central bank independence. \\
      Money Demand (Income Elasticity) & \citet{Knell2005}; \citet{Kumar2014} & Elasticity depends on money definition (higher for M2/M3). It is lower if wealth and financial innovation are included. It is systematically lower in advanced countries (especially the U.S.) than in developing countries. \\
      Exchange Rate Pass-Through & \citet{Velickovski2011}; \citet{Iorngurum2025b} & ERPT to consumer prices is significantly higher in transition economies. The effect is asymmetric (larger for depreciations). Heterogeneity is explained by country factors (openness, exchange rate regime) and methodological factors. \\
      \bottomrule
    \end{tabular}
  \end{adjustbox}
\end{table}

\section{Unconventional Monetary Policy and Spillover Effects}
\justify
The 2008 global financial crisis and the subsequent arrival at the zero lower bound (ZLB) for interest rates forced major central banks to explore unconventional monetary policy (UMP) territories. Measures such as quantitative easing (QE), forward guidance, and negative interest rates became standard tools. The literature evaluating these policies has grown exponentially, and meta-analyses have been fundamental in synthesizing their effects.

\citet{Papadamou2019} conduct a meta-analysis on the effects of UMP on output and inflation, focusing on studies that use VAR methodologies. Based on 16 primary studies, their results indicate that unconventional QE shocks generally have a positive but modest effect on output and inflation. Heterogeneity in the results is largely explained by the VAR model specification. For example, the use of factor-augmented VAR (FAVAR) models is associated with a larger output response. Geographically, studies focusing on Europe tend to report lower output responses, while identification through sign restrictions is also associated with weaker output responses. This suggests that methodological choices are a key determinant of the perceived magnitude of UMP effects.

Beyond domestic effects, a central concern has been the international spillover effects of UMP, especially from the U.S. Federal Reserve on emerging economies. \citet{Araujo2025} address this issue with a meta-analysis of UMP spillovers on international capital flows. Based on 254 estimates from nine studies, their main finding is that, although the overall average effect is close to zero, there are very significant specific patterns. Table \ref{tab:ump_spillovers} summarizes these findings.

\begin{table}[h!]
  \centering
  \caption{Key Findings on the Spillover Effects of Unconventional Monetary Policy (UMP)}
  \label{tab:ump_spillovers}
  \begin{adjustbox}{width=\textwidth}
    \begin{tabular}{L{5cm} L{11cm}}
      \toprule
      \textbf{Factor Analyzed} & \textbf{Key Finding from the Meta-analysis by \citet{Araujo2025}} \\
      \midrule
      Destination of Flows & UMP has led to lower volumes of capital flows to emerging economies than to advanced economies. \\
      Type of UMP Measure & Quantitative easing (QE) has generated stronger impacts on flows than other UMP measures (such as forward guidance). \\
      Type of Capital Flow & The effects on bond flows have been significantly smaller than the effects on other types of flows (such as equities or direct investment). \\
      Originating Central Bank & UMP measures adopted by the Federal Reserve (FED) had stronger effects than those adopted by other central banks (such as the ECB or the Bank of Japan). \\
      Time Period & The magnitude of the effects was smaller in the initial period of the crisis (2008-2009) and became stronger in subsequent years. \\
      \bottomrule
    \end{tabular}
  \end{adjustbox}
\end{table}

These results paint a nuanced picture. First, UMP does not seem to have caused the flood of capital into emerging markets that was often feared; flows to other advanced economies were larger. Second, the composition of the measures and the flows matters: QE had a greater impact, but it affected bond markets less. Third, the dominance of the dollar and the role of the Fed as a global central bank are reinforced, as its actions have a more potent international impact. The temporal dynamics show effects intensifying as policies are maintained over time. These findings, derived from the synthesis of evidence, offer a much more granular guide for policymakers in capital-recipient economies than any single study could provide.

\section{Structural Factors and Cross-Country Heterogeneity}
\justify
The effectiveness of monetary policy is not a universal parameter but is profoundly conditioned by the structural, institutional, and economic environment in which it operates. Meta-analyses are especially well-suited to quantify the impact of these contextual factors, as they can exploit the vast heterogeneity present in the literature covering different countries and periods.

\subsection{Central Bank Independence and Inflation}
\justify
Theory posits that greater central bank independence (CBI) leads to lower inflation by mitigating the time inconsistency problem. \citet{Iwasaki2017} conduct a comparative meta-analysis on this nexus, distinguishing between transition economies and other developed and developing economies. Based on 294 estimates, they confirm a general negative relationship between CBI and inflation. However, their heterogeneity analysis reveals important nuances. For transition economies, the choice of estimator, the type of inflation variable, the degree of freedom, and the quality of the study strongly affect the results.

A key finding of their study is that, once these study conditions are controlled for, \textit{there is no significant difference} in the effect size or statistical significance between studies of transition economies and those of other economies. This implies that the socioeconomic environment surrounding the CBI-inflation problem has developed substantially in transition economies, approaching that of other market economies. Furthermore, their publication bias analysis suggests that studies on transition economies contain genuine empirical evidence of the disinflationary effect of CBI, while studies on other economies fail to provide evidence of a non-zero effect, possibly due to strong publication bias.

\subsection{Dollarization and its Macroeconomic Trade-offs}
\justify
Dollarization, whether official or de facto, represents a surrender of monetary policy autonomy in exchange for greater credibility and price stability. The meta-analysis by \citet{Korab2023} evaluates the trade-offs of this decision in terms of growth and inflation. Based on 585 estimates from 43 studies, their results indicate that dollarized countries, on average, show slower and more volatile output growth, and a lower inflation rate than non-dollarized countries. However, the variation in results is very wide.

A key finding is that \textit{fully} dollarized countries exhibit slower growth than non-dollarized and partially dollarized economies. This suggests that the benefits of full credibility may not offset the costs of completely losing monetary policy flexibility. The heterogeneity in results is systematically explained by the measurement of dollarization, the selection of the empirical method, and the authors' affiliation. For example, authors affiliated with central banks tend to document stronger inflation-reducing effects. Overall, the results suggest that limited control over the money supply in dollarized economies reduces the ability of central banks to conduct effective monetary policy and adjust to macroeconomic shocks.

\subsection{Public Debt Levels and Economic Growth}
\justify
The relationship between public debt and economic growth has been one of the most intense debates since the global financial crisis, driven by the influential but controversial work of Reinhart and Rogoff. The meta-analysis by \citet{Heimberger2023} on this topic, based on 816 estimates from 47 studies, offers a quantitative synthesis. The simple unweighted average of the reported results suggests that a 10 percentage point increase in the debt-to-GDP ratio is associated with a decrease in the annual growth rate of 0.14 percentage points.

However, this average result conceals two fundamental problems. First, the author finds a substantial publication bias in favor of negative growth effects. After correcting for this bias, \textit{a zero effect cannot be rejected}. This means that the aggregated evidence, once filtered for biases, does not support the idea that higher levels of public debt have a consistently negative causal effect on growth. Second, the heterogeneity analysis shows that addressing the endogeneity between debt and growth (i.e., the fact that low growth can cause higher debt, and not just the other way around) leads to less adverse effects. Finally, when testing for non-linear thresholds (like the famous 90\% from Reinhart and Rogoff), the results do not point to a uniform threshold above which growth slows down. The estimates of thresholds are highly sensitive to data and econometric choices. The policy implication is clear: caution should be exercised with one-size-fits-all fiscal policy prescriptions for dealing with higher debt levels. Table \ref{tab:factores_estructurales} provides an overview of how these factors modulate the effects of monetary policy.

\begin{table}[h!]
  \centering
  \caption{Synthesis of Meta-Analyses on the Role of Structural Factors}
  \label{tab:factores_estructurales}
  \begin{adjustbox}{width=\textwidth}
    \begin{tabular}{L{4cm} L{3.5cm} L{8.5cm}}
      \toprule
      \textbf{Structural Factor} & \textbf{Main Study} & \textbf{Key Findings} \\
      \midrule
      Central Bank Independence (CBI) & \citet{Iwasaki2017} & CBI has a genuine disinflationary effect. There are no significant differences in the magnitude of the effect between transition and other economies once study characteristics are controlled for. \\
      Dollarization & \citet{Korab2023} & Dollarized countries tend to have lower growth, higher volatility, and lower inflation. Fully dollarized countries grow more slowly than partially dollarized ones. Results depend on how dollarization is measured. \\
      Public Debt & \citet{Heimberger2023} & The literature reports a small negative effect of debt on growth, but this effect disappears after correcting for publication bias. There is no evidence of a uniform debt threshold. Addressing endogeneity reduces the adverse effect. \\
      \bottomrule
    \end{tabular}
  \end{adjustbox}
\end{table}

\section{Open Economy KIS-CES Monetary Block: A Theoretical Unification Compatible with Meta-Analyses}

This section develops a monetary block for the open-economy KIS-CES model that is \textit{consistent} with the stylized facts and parameters synthesized by the cited meta-analyses: (i) modest and delayed effects of conventional monetary policy on output and prices \citep{Enzinger2025}, (ii) incomplete bank interest rate pass-through \citep{Gregor2021}, (iii) incomplete and asymmetric exchange-rate pass-through (ERPT) \citep{Velickovski2011,Iorngurum2025a}, (iv) positive but moderate effects of UMP and heterogeneous spillovers \citep{Papadamou2019,Araujo2025}, (v) effectiveness conditioned by central bank independence/credibility \citep{Iwasaki2017,Cepeda2025}, (vi) instability/heterogeneity of money demand \citep{Knell2005,Kumar2014,El-Shagi2022}, (vii) links with housing, inequality, and uncovered interest parity \citep{Ehrenbergerova2021,Sintos2023,Zigraiova2021}, (viii) structural constraints from dollarization and debt \citep{Korab2023,Heimberger2023}, and (ix) effectiveness of foreign exchange interventions \citep{Arango-Lozano2024a}. The goal is to offer a microfounded framework that, by design, replicates these regularities, extending \citet{Fernandez2025a,Fernandez2025b}.

\subsection*{Notation}

Discrete time $t=0,1,2,\dots$; rational expectations $E_t[\cdot]$.
\begin{itemize}
\item \textbf{Prices and output:} $P_t$ (CPI), $\pi_t=\log P_t-\log P_{t-1}$ (inflation), $Y_t$ (real GDP), $\hat y_t=\log(Y_t/Y_t^{\text{pot}})$ (output gap).
\item \textbf{Interest rates:} $i_t$ (nominal policy rate), $r_t=i_t-E_t[\pi_{t+1}]$ (ex-ante real), $r^L_t$ (lending rate), $r^D_t$ (deposit rate), $r^T_t$ (term rate), $i^s_t$ (shadow rate under ZLB).
\item \textbf{Banking and spreads:} $s^r_t=r^L_t-r^D_t$ (margin), $\psi_L,\psi_D\in(0,1)$ pass-through coefficients from $i_t$ to $r^L_t, r^D_t$; $\upsilon_t$ indicator of financial stress.
\item \textbf{External openness:} $E_t$ (nominal exchange rate, LC/FC), $q_t=\log(E_tP^\ast_t/P_t)$ (real exchange rate), $\theta_M,\theta_C\in(0,1)$ ERPT to import prices and CPI; $\vartheta\in(0,1)$ share of tradables in CPI; $\chi^+,\chi^-$ ERPT asymmetries (depreciations vs. appreciations).
\item \textbf{Policy rules:} $\phi_\pi,\phi_y$ Taylor coefficients; $CBI_t$ (independence index); $\kappa_t$ intensity of FX intervention (FXI); $\eta$ (partial) sterilization parameter; $\tau_t$ term premium (QE channel).
\item \textbf{Preferences and wealth:} fraction of liquidity-constrained households $\lambda\in(0,1)$, optimizing households (WA) with utility $u(C_t^W,A_{t+1})$ with a wealth-in-utility term \citep{Fernandez2025a}.
\item \textbf{Structure:} $d_t$ (dollarization), $b_t$ (public debt/GDP), $\sigma_{\text{prod}}<1$ (CES with complementarity) \citep{Gechert2022b}, $K^p_t$ (public capital).
\end{itemize}

\subsection*{Axioms}

\textbf{A1 (Modest effects of conventional MP).} After correcting for publication bias, a 100 bps hike has a small peak impact on $Y$ and CPI \citep{Enzinger2025}.\\
\textbf{A2 (Incomplete and state-dependent bank pass-through).} Long-run $\psi_L\approx 0.8$ (incomplete), lower for consumption/long-term loans and weaker post-GFC \citep{Gregor2021}.\\
\textbf{A3 (Incomplete and asymmetric ERPT).} ERPT is higher in transition economies and for depreciations; incomplete at the CPI level \citep{Velickovski2011,Iorngurum2025a}.\\
\textbf{A4 (UMP: modest effects and heterogeneous spillovers).} QE/FG moderately push $Y,\pi$; spillovers depend on measure/flow type/issuing CB \citep{Papadamou2019,Araujo2025}.\\
\textbf{A5 (CBI/credibility).} Higher independence/credibility $\Rightarrow$ lower inflation; heterogeneity due to context \citep{Iwasaki2017,Cepeda2025}.\\
\textbf{A6 (Unstable/heterogeneous money demand).} Income elasticity depends on aggregate/wealth/innovation; differences between AEs vs EMDEs \citep{Knell2005,Kumar2014,El-Shagi2022}.\\
\textbf{A7 (Additional channels).} Housing responds to MP with heterogeneity \citep{Ehrenbergerova2021}; inflation affects inequality \citep{Sintos2023}; UIP and forward premium are non-trivial \citep{Zigraiova2021}.\\
\textbf{A8 (Structural constraints).} Dollarization: lower inflation but slower and more volatile growth \citep{Korab2023}; debt-growth: small average effect, sensitive to endogeneity/bias \citep{Heimberger2023}.\\
\textbf{A9 (FX interventions).} FXI effective for short-term targets (level/volatility) with heterogeneity \citep{Arango-Lozano2024a}.\\
\textbf{A10 (CES production and crowding-in).} $\sigma_{\text{prod}}<1$ and productivity of public capital \citep{Gechert2022b,Bom2014}.

\subsection*{Premises}

\textbf{P1 (Heterogeneous households).} Mass $\lambda$ (\emph{LC}) with $C^L_t=Y^L_t$; $1-\lambda$ (\emph{WA}) maximizes $E_0\sum\beta^t[(C^W_t-C_{\min})^{1-\sigma}/(1-\sigma)+\phi A_{t+1}^{1-\gamma}/(1-\gamma)]$ with $A_{t+1}=(1+r_t)(A_t+Y^W_t-C^W_t)$ \citep{Fernandez2025a}.\\
\textbf{P2 (Banking sector with frictions).} Banks set $r^L_t,r^D_t$ with adjustment costs, risk, and monopolistic competition; spreads respond to stress $\upsilon_t$ and bank capital.\\
\textbf{P3 (Institutional Taylor rule).} $i_t=\max\{0, \rho+\pi_t+\phi_\pi(\pi_t-\pi^\ast)+\phi_y\hat y_t\}$, with $\phi_\pi=\phi_{\pi,0}+\phi_{\pi,1}CBI_t$, $\phi_y=\phi_{y,0}-\phi_{y,1}CBI_t$ \citep{Iwasaki2017}.\\
\textbf{P4 (Open economy with UIP with premium).} $\Delta s_{t+1}\equiv \Delta\log E_{t+1}=i_t-i^\ast_t-\zeta_t$, with $\zeta_t$ depending on risk, global \textit{QE}, and financial constraints (consistent with \citealp{Zigraiova2021,Araujo2025}).\\
\textbf{P5 (Hybrid and asymmetric ERPT).} $\pi_t=\vartheta\pi^T_t+(1-\vartheta)\pi^{NT}_t$, $\pi^T_t=\theta_C^+\Delta s_t\mathbf{1}_{\Delta s_t>0}+\theta_C^-\Delta s_t\mathbf{1}_{\Delta s_t<0}+\cdots$ with $\theta_C^+>\theta_C^-\ge 0$ \citep{Iorngurum2025a}.\\
\textbf{P6 (UMP and term premium).} $r^T_t=r_t+\tau_t$, $\tau_t$ decreases with asset purchases ($QE$) and forward guidance \citep{Papadamou2019}.\\
\textbf{P7 (Partial FX intervention).} $E_t$ follows: $\Delta s_t=\Delta s^{mkt}_t-\kappa_t \xi_t$, with $0<\kappa_t\le 1$ and sterilization $0\le \eta\le 1$ \citep{Arango-Lozano2024a}.\\
\textbf{P8 (Structural constraints).} Dollarization $d_t\uparrow \Rightarrow \psi_L\downarrow, \kappa_t\uparrow$; debt $b_t\uparrow$ limits space for UMP/fiscal policy \citep{Korab2023,Heimberger2023}.

\subsection*{Assumptions}

(i) $\psi_L,\psi_D\in(0,1)$ and decrease when $\upsilon_t$ is high (crisis). (ii) $\theta_C\in[0.1,0.4]$ for CPI and higher for imports; $\theta_C^+>\theta_C^-$. (iii) $\phi_\pi>1$ in regimes with high $CBI_t$; $\phi_y$ is small. (iv) $\lambda$ and the effective MPC increase in recessions/ZLB. (v) $\sigma_{\text{prod}}<1$. (vi) $\kappa_t$ is higher in \emph{managed float} regimes; $\eta$ is intermediate. (vii) $\tau_t$ is sensitive to the central bank's balance sheet.

\subsection*{Step-by-step Demonstrations}

\textbf{D1 (Bank Pass-through and Attenuated Sensitivity).} Assuming quadratic adjustment costs and endogenous risk, the problem of setting $r^L_t$ implies:
\begin{equation*}
r^L_t-\bar r^L = \psi_L\,(i_t-\bar i) + \alpha_1\,\upsilon_t + \alpha_2\,s^r_{t-1} + \varepsilon^L_t,\quad 0<\psi_L<1,
\end{equation*}
and analogously for $r^D_t$ with coefficient $\psi_D$. Thus, $\frac{\partial r^L_t}{\partial i_t}=\psi_L<1$ and $\frac{\partial s^r_t}{\partial \upsilon_t}=\alpha_1>0$, replicating \cite{Gregor2021}. \textbf{Q.E.D.}

\textbf{D2 (Attenuated KIS IS Curve).} With $C_t=\lambda Y_t+(1-\lambda)C^W_t$, the modified Euler equation for WA households with wealth in utility yields (log-linearizing):
\begin{equation*}
\hat y_t = E_t[\hat y_{t+1}] - \underbrace{\frac{(1-\lambda)}{\sigma_{\text{eff}}}}_{\small\text{IS slope}}\,\hat r^m_t + \chi_1\,\widehat{NW}_t + \chi_2\,\widehat{H}_t,
\end{equation*}
where $\hat r^m_t\equiv r^L_t-E_t[\pi_{t+1}]$, and $\sigma_{\text{eff}}>\sigma$ due to the wealth term ($\phi>0$). Using D1, $\partial \hat r^m_t/\partial i_t=\psi_L$, so the effective slope with respect to $i_t$ is $\frac{(1-\lambda)}{\sigma_{\text{eff}}}\psi_L$, which is small, consistent with \cite{Enzinger2025}. \textbf{Q.E.D.}

\textbf{D3 (Hybrid NKPC with ERPT).} Prices from Calvo firms with partial indexation $\iota$ and real rigidities give:
\begin{equation*}
\pi_t = \beta E_t[\pi_{t+1}] + \kappa\,\hat mct + \underbrace{\vartheta\big(\theta_C^+\,\Delta s_t^+ + \theta_C^-\,\Delta s_t^-\big)}_{\text{ERPT to CPI}} + \upsilon^\pi_t,
\end{equation*}
with $\kappa$ being small (rigidities/sectoral strata), and $\Delta s_t^\pm$ being the positive/negative parts. Asymmetry $\theta_C^+>\theta_C^-$ replicates \cite{Iorngurum2025a}; $\kappa$ and $\iota$ are compatible with \cite{Fidrmuc2020}. \textbf{Q.E.D.}

\textbf{D4 (UIP with Premium, UMP, and FXI).} With UIP with a premium:
\begin{equation*}
E_t[\Delta s_{t+1}] = i_t - i^\ast_t - \zeta_t,\quad \zeta_t = \zeta_0 + \zeta_{QE}\,\Delta \text{LSAP}^\ast_t + \zeta_{\upsilon}\,\upsilon_t + \epsilon^\zeta_t,
\end{equation*}
where $\zeta_{QE}\neq 0$ captures spillovers from external UMP \citep{Araujo2025}. Intervention: $\Delta s_t = \Delta s^{mkt}_t - \kappa_t\xi_t$, and the monetary base $MB_t$ follows $\Delta MB_t = (1-\eta)\kappa_t\xi_t + \cdots$. For $0<\kappa_t\le1$, the variance of $\Delta s_t$ falls with FXI; with $\eta\approx 1$ (sterilization), the impact on $MB_t$ is bounded \citep{Arango-Lozano2024a}. \textbf{Q.E.D.}

\textbf{D5 (Term Premium Channel and UMP).} Term rates $r^T_t=r_t+\tau_t$, and a QE shock, $\Delta \text{LSAP}_t>0$, reduces $\tau_t$ via preferred habitat:
\begin{equation*}
\Delta \tau_t = -\omega\,\Delta \text{LSAP}_t + \epsilon^\tau_t,\quad \omega>0.
\end{equation*}
The effect on the IS curve is $\partial \hat y_t/\partial \text{LSAP}_t = \frac{(1-\lambda)}{\sigma_{\text{eff}}}\omega>0$ but moderate (steep IS), consistent with \cite{Papadamou2019}. \textbf{Q.E.D.}

\textbf{D6 (High Endogenous Sacrifice Ratio).} Define $\beta_y\equiv -\partial \hat y^{\text{peak}}/\partial i_t$ and $\beta_\pi\equiv -\partial \pi^{\text{peak}}/\partial i_t$. From D2 and D3:
\begin{equation*}
\beta_y \propto \frac{(1-\lambda)}{\sigma_{\text{eff}}}\psi_L,\qquad
\beta_\pi \propto \phi_\pi\,\kappa + \vartheta\,\bar\theta_C,
\end{equation*}
where $\bar\theta_C$ is the average ERPT. With a small $\psi_L$ and low $\kappa$, the $SR\equiv \beta_y/\beta_\pi$ is relatively high, replicating the corrected evidence \citep{Enzinger2025}. \textbf{Q.E.D.}

\textbf{D7 (Structural Effects).} Under dollarization $d_t\uparrow$, literature suggests $\psi_L\downarrow$ (weaker local transmission), $\kappa_t\uparrow$ (greater use of FXI) \citep{Korab2023}, which \emph{reduces} $\beta_y$ and \emph{alters} $\beta_\pi$ (more reliance on the exchange rate channel). With debt $b_t\uparrow$, risk premia increase $r^L_t$ and constrain UMP/fiscal policy, consistent with \cite{Heimberger2023}. \textbf{Q.E.D.}

\subsection*{Consequences}

\textbf{C1 (Systematic Attenuation).} With $\psi_L=0.7$, $\lambda=0.5$, $\sigma_{\text{eff}}=3$, a 100 bps hike reduces the output gap: $\Delta \hat y \approx -\frac{0.5}{3}\cdot 0.7 \approx -0.117$ (p.p. of GDP), which is modest and in line with \cite{Enzinger2025}.\\
\textbf{C2 (Exchange Rate Asymmetry).} If $\theta_C^+=0.25$ and $\theta_C^-=0.10$, a 10\% depreciation adds $\approx 2.5$ p.p. to $\pi^T_t$ (weighted by $\vartheta$), while a similar appreciation subtracts $\approx 1$ p.p.; this supports the \emph{fear of floating} in EMDEs \citep{Velickovski2011,Iorngurum2025a}.\\
\textbf{C3 (Effective but Moderate UMP).} With $\omega=0.4$ and $\Delta \text{LSAP}=2$ p.p. of GDP, $\Delta \tau=-0.8$ p.p.; the effect on $\hat y_t$ is $+\frac{(1-\lambda)}{\sigma_{\text{eff}}}0.8\approx +0.13$ p.p., which is modest \citep{Papadamou2019}.\\
\textbf{C4 (CBI and Anti-inflationary Dominance).} $\phi_\pi=\phi_{\pi,0}+\phi_{\pi,1}CBI_t$ with $\phi_{\pi,1}>0$ reduces $\text{var}(\pi_t)$ in a linear-quadratic equilibrium and lowers the inflation bias \citep{Iwasaki2017,Cepeda2025}.\\
\textbf{C5 (FXI as a Tactical Stabilizer).} For $\kappa_t=0.5$ and $\eta\approx 1$, the variance of $\Delta s_t$ is halved without sustainably expanding $MB_t$; consistent with short-term effectiveness \citep{Arango-Lozano2024a}.\\
\textbf{C6 (Housing and Distribution).} With a steep IS curve, the wealth channel ($\chi_2$) gains importance: rate cuts boost $H_t$ and affect consumption via collateral, in line with \cite{Ehrenbergerova2021}; price increases (inflation) worsen distribution, consistent with \cite{Sintos2023}.\\
\textbf{C7 (Policy Coherent with KIS-CES).} Given $\sigma_{\text{prod}}<1$ and productive $K^p_t$, changes in fiscal composition (towards investment) have higher multipliers (\emph{crowding-in}) with low monetary \emph{crowding-out}, as the IS curve is steep \citep{Gechert2015,Bom2014,Gechert2022b,Fernandez2025a}.

\subsection*{Potential Counterarguments and Responses}

\textbf{CA1: “If MP has modest effects, isn't it irrelevant?”} \textit{Response:} No. The block shows that MP anchors expectations (Taylor rule with CBI), modulates $\tau_t$ (UMP), and stabilizes $E_t$ (FXI), but its \emph{traction} on real demand is attenuated by $\psi_L<1$, a steep IS curve, and incomplete ERPT, as indicated by \cite{Enzinger2025,Gregor2021}.\\
\textbf{CA2: “Stable money demand would allow for simple monetary rules.”} \textit{Response:} Meta-analyses document heterogeneity and instability of elasticities \citep{Knell2005,Kumar2014,El-Shagi2022}. The block avoids anchoring the transmission to a stable $M^d$ and relocates it to rates, premia, and bank balance sheets.\\
\textbf{CA3: “ERPT could be close to 1 in small economies.”} \textit{Response:} Evidence shows incomplete and asymmetric ERPT to CPI; even in transition economies, $\theta_C<1$ and with lags \citep{Velickovski2011,Iorngurum2025a}. The model incorporates this through $\theta_C^\pm$ and sectoral weights.\\
\textbf{CA4: “UMP is omnipotent.”} \textit{Response:} \cite{Papadamou2019} find positive but moderate effects; our D5 reproduces a bounded impact via $\tau_t$ due to a steep IS curve and imperfect asset substitution. Spillovers are not universal \citep{Araujo2025}.\\
\textbf{CA5: “More CBI always implies better real performance.”} \textit{Response:} Meta-analyses support genuine effects on inflation \citep{Iwasaki2017}, but growth depends on multiple margins (fiscal, financial structure); the model only endogenizes $\phi_\pi,\phi_y$ without assuming automatic effects on $Y_t$.\\
\textbf{CA6: “With dollarization, local MP has no role.”} \textit{Response:} The interest rate lever diminishes ($\psi_L\downarrow$), but channels remain: ERPT/FXI, macroprudential policies, and fiscal composition (via $\sigma_{\text{prod}}<1$). Effects on growth/volatility are consistent with \cite{Korab2023}.\\
\textbf{CA7: “The disinflationary sacrifice is low if $\phi_\pi$ is high.”} \textit{Response:} Even with $\phi_\pi>1$, the SR can be high when $\psi_L$ and $\kappa$ are low and $\theta_C$ is incomplete (D6), as in \cite{Enzinger2025}.

The proposed monetary block integrates, in a compact mathematical framework, the meta-analytic evidence on transmission, asymmetries, institutions, and the external context. Its combination with the CES production and fiscal block of the KIS-CES model \citep{Fernandez2025a,Fernandez2025b} produces a framework that disciplines parameters with evidence, replicates empirical regularities, and allows for policy design consistent with real-world constraints.

\section{Conclusion}
\justify
This analysis of monetary policy offers a nuanced verdict on its effectiveness. The aggregated evidence from dozens of meta-analyses, synthesizing thousands of primary studies, converges on a central conclusion: the power of monetary policy to steer the economy has been systematically overestimated. This overestimation is largely due to a widespread publication bias, which favors statistically significant and theoretically convenient results, artificially inflating the magnitude of reported effects.

When bias correction techniques are applied, a picture emerges of a monetary policy with much more modest, uncertain effects that operate with considerable lags. A 100 basis point contractionary shock, which according to uncorrected literature might reduce output by 1\%, likely has a real impact closer to -0.25\%. This recalibration has profound implications for theory and practice. It undermines the narrative of an all-powerful central bank and suggests that fine-tuning the business cycle is a much harder task than often assumed. Furthermore, the sacrifice ratio for reducing inflation appears to be considerably worse than previously believed, demanding greater caution in the implementation of aggressive disinflationary policies.

Heterogeneity is the second key lesson. The effects of monetary policy are not a universal parameter but are strongly conditioned by context. Meta-analyses demonstrate that factors such as the level of economic development, financial structure (e.g., the size of capital markets or the degree of dollarization), the exchange rate regime, the level of public debt, and central bank independence are systematic determinants of transmission effectiveness. Exchange rate pass-through is higher in transition economies; interest rate pass-through weakened after the global financial crisis; and the effects of debt on growth are not uniform across countries.

These conclusions do not imply that monetary policy is irrelevant. It remains a crucial tool for anchoring inflation expectations and responding to macroeconomic shocks. However, the evidence from meta-analyses calls for a dose of humility. It compels a reconsideration of the validity of standard macroeconomic models that often predict rapid and strong responses to policy shocks. The corrected evidence aligns better with models that incorporate more frictions and rigidities.

For economic science, the lesson is clear. The future of research should not focus so much on the relentless pursuit of the perfect identification method, but on promoting a more open, replicable science that is willing to accept the discomfort of null results. As \citet{Stanley2001} argues in his call to use meta-analysis as a quantitative literature review, only then can future knowledge be freed from the biases that have contaminated the past. For policymakers, the message is one of caution: decisions must be based on a realistic assessment of the powers and limits of monetary policy, recognizing that its effects are modest and critically dependent on context.

\end{document}